\documentclass[twocolumn,aps,superscriptaddress,preprintnumbers]{revtex4-1}

\usepackage{amsmath,amssymb}
\usepackage{graphics}
\usepackage{graphicx}
\usepackage{dcolumn}
\usepackage{bm}
\usepackage{soul}
\usepackage{multirow}
\usepackage{setspace}
\usepackage[mathlines]{lineno}
\usepackage{textcomp}
\usepackage{mathcomp}
\usepackage{spverbatim}
\usepackage{float}
\usepackage{wasysym}

\usepackage{xcolor}

\usepackage{hyperref}
\hypersetup{
    colorlinks=true,       
    linkcolor=blue,        
    citecolor=blue,        
    filecolor=blue,        
    urlcolor=blue,         
    runcolor=blue
}

\usepackage{enumitem,amssymb}
\newlist{todolist}{itemize}{2}
\setlist[todolist]{label=$\square$}
\usepackage{pifont}

\newcommand{\ket}[1]{\left| #1 \right\rangle}
\newcommand{\bra}[1]{\left\langle #1 \right|}

\newcommand{\average}[3]{\left\langle #1 \right| #2 \left| #3 \right\rangle}
\newcommand{\meanval}[1]{\left\langle #1 \right\rangle}

\newcommand{\ic}{\dot{\imath}}

\newcommand{\USACH}{Department of Physics, Universidad de Santiago de Chile, Av. Victor Jara 3493, Santiago, Chile}
\newcommand{\MIRO}{ANID-Millennium Institute for Research in Optics, Chile}

\begin{document}

\title{Open quantum dynamics of strongly coupled oscillators with multi-configuration time-dependent Hartree propagation and Markovian quantum jumps}

\author{Johan F. Triana}
\affiliation{\USACH}

\author{Felipe Herrera}
\affiliation{\USACH}
\affiliation{\MIRO}

\date{\today}

\begin{abstract}
Modeling the non-equilibrium dissipative dynamics of strongly interacting quantized degrees of freedom is a fundamental problem in several branches of physics and chemistry. We implement a quantum state trajectory scheme for solving Lindblad quantum master equations that describe coherent and dissipative processes for a set of strongly-coupled quantized oscillators. The scheme involves a sequence of stochastic quantum jumps with transition probabilities determined the system state and the system-reservoir dynamics. Between consecutive jumps, the wavefunction is propagated in coordinate space using the multi-configuration time-dependent Hartree (MCTDH) method. We compare this hybrid propagation methodology with exact Liouville space solutions for physical systems of interest in cavity quantum electrodynamics, demonstrating accurate results for experimentally relevant observables using a tractable number of quantum trajectories. We show the potential for solving the dissipative dynamics of finite size arrays of strongly interacting quantized oscillators with high excitation densities, a scenario that is challenging for conventional density matrix propagators due to the large dimensionality of the underlying Hilbert space.  
\end{abstract}

\maketitle

\section{\label{sec:intro}Introduction}

Accurate numerical simulations of open quantum systems are fundamentally important for the development of quantum technology \cite{Koch_2016}. Understanding and possibly controlling system-reservoir interactions enables a diverse set of applications such as the manipulation of quantum speed limits for driven state evolution \cite{Deffner2013}, quantum metrology with improved precision bounds \cite{Chin2012,Haase2016}, quantum circuits with improved gate fidelities  \cite{Di-Paolo2021}, quantum optics with nanophotonics \cite{Tame2013,Schmidt2016}, or controlled chemistry with quantum optics \cite{Herrera2016,Herrera2020perspective}. In many applications, the  temporal correlations of the reservoir variables that couple with the system of interest decay much faster than the system-reservoir interaction times. The open quantum system dynamics can then be  modeled with Markovian quantum master equations for the evolution of the reduced system density matrix $\hat\rho_S$ \cite{Breuer2002,Manzano2020}. For a Hilbert space of dimension $d$, the density matrix scales with $d^2$, making the direct integration of quantum master equations numerically intractable for large many-body problems, as $d$ scales exponentially with the number of particles \cite{Weimer2021}.

To simulate the dynamics of open quantum many-body systems, several techniques have been developed, including stochastic methods \cite{Gisin1992,Dalibard1992,Molmer93}, tensor networks representations \cite{Verstraete2004,Werner2016,Orus2019}, phase space methods \cite{Carusotto2005,Navez2010,Schachenmayer2015}, variational methods \cite{Weimer2015}, cluster expansions \cite{Cao1990,Tanimura2020,TANG2013} and field theory techniques \cite{Cao1994,Cao1994II,Sieberer2016}. Broadly speaking, these approaches differ in the way the density matrix and quantum master equation are represented and propagated. Advanced simulation techniques are commonly used in chemical physics for treating strongly molecules in complex reservoirs \cite{Valleau2012,Moix2013,delPino2018,Wang2020}. Cavity quantum electrodynamics (QED) with molecules has emerged as another domain in which advanced quantum dynamics methods are useful \cite{GarciaVidal2021,Feist2018,Simpkins2021,Herrera2020perspective}, as the emergence of cavity-induced single-particle and many-body  correlations is believed to be relevant for explaining experiments on the modification of chemical reaction rates in optical and infrared cavities  \cite{Herrera2016,Fregoni2019,Felicetti2020,Fregoni2020,Campos-Gonzalez-Angulo2019,Foley2020,Du2018,delPino2018,Ulusoy2020,Gu2020,Ahn2022}.

We develop a stochastic wavefunction methodology for solving Markovian quantum master equations in the coordinate representation. The stochastic component of the method is based on the Dalibard-Castin-Molmer quantum jump technique \cite{Molmer92,*Molmer93}, a type of Monte Carlo method \cite{Plenio1998} where the system wavefunction undergoes a sequence of quantum jumps with transition probabilities determined the instantaneous state of the system and the physics of the system-reservoir coupling. Between consecutive quantum jumps the wavefunction evolves deterministically according to the instantaneous Hamiltonian, which we represent in coordinate space using the multi-configuration time-dependent Hartree (MCTDH) method \cite{Meyer1990,*Beck2000,MeyerMCTDH2007}. Observables in the quantum jump technique are guaranteed to converge to the density matrix solution of the quantum master equation by averaging over a sufficient number of wavefunction trajectories \cite{*Molmer93}. The scheme is applicable to Markovian master equations in Lindblad form \cite{Manzano2020}, but extensions to more general reservoirs have been developed \cite{deVega2017,Piilo2009}.

Wavefunction trajectories in the coordinate representation are particularly well suited for studying  strongly interacting oscillators subject to driving and dissipation, as often found in molecular cavity QED problems \cite{Feist2018,Herrera2020perspective,GarciaVidal2021}. MCTDH  propagators can already capture  strong correlations between high-dimensional anharmonic oscillators that naturally emerge in chemical physics \cite{Raab2000,Nest2003,Andrianov2006,Vendrell2011,VENDRELL2018}. Therefore, extending the MCTDH method beyond the use of complex potentials \cite{Ulusoy2020} is a significant step toward  scalable atomistic modeling of many-body molecular cavity QED systems.

In what follows, we briefly review the quantum jump and MCTDH methods (Sec. \ref{sec:mcwfmctdh}). Then demonstrate the applicability of the proposed methodology and explore its limitations (Sec. \ref{sec:results}) and suggest possible applications of the method for studying cavity QED with molecular oscillators (Sec. \ref{sec:conclu}).

\section{Methods}
\label{sec:mcwfmctdh}


\subsection{Monte Carlo Wavefunction Method}
\label{sec:mcmctdh}

 \begin{figure}[!t]
 \includegraphics[width=0.45\textwidth]{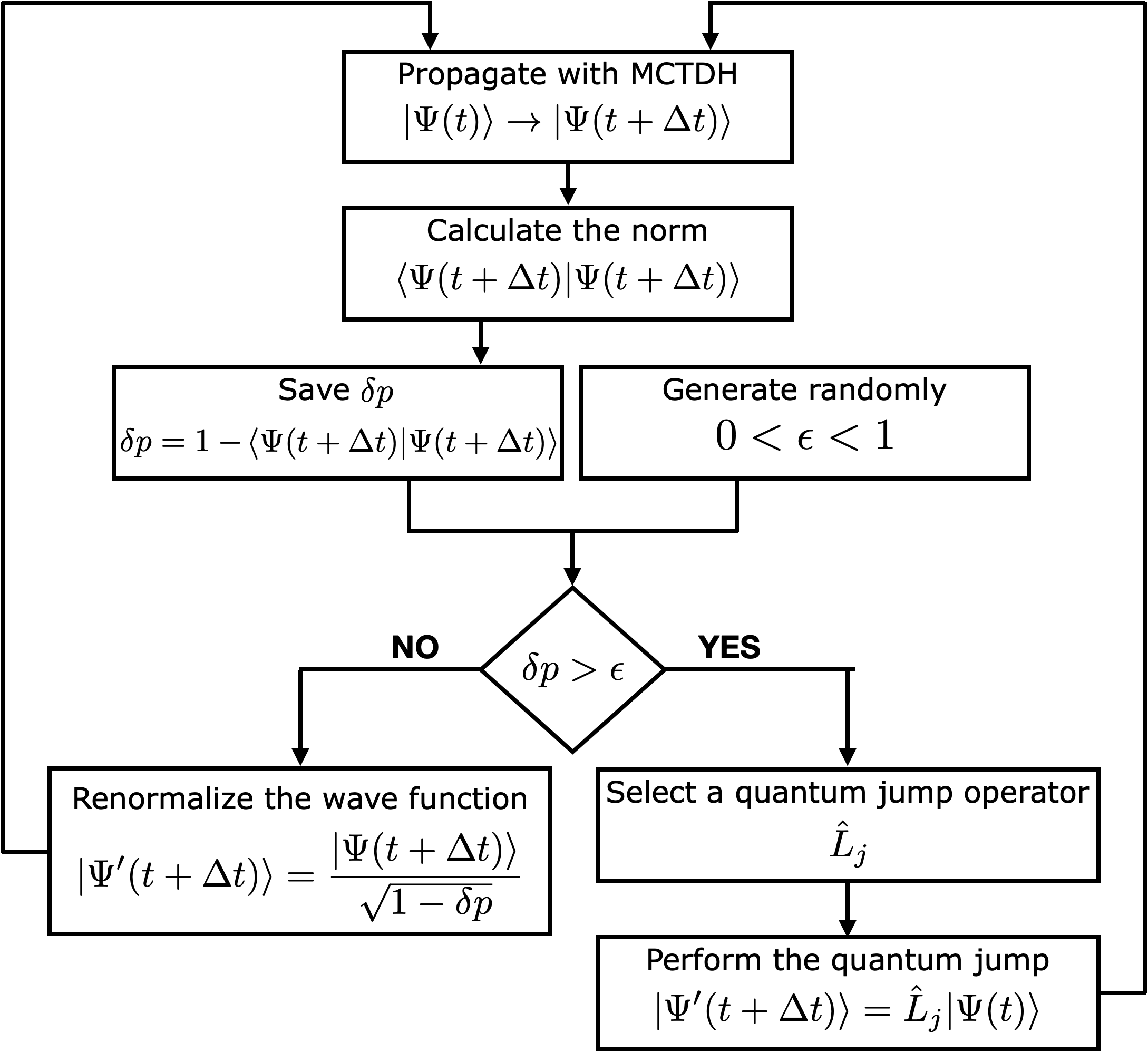}
 \caption{{\bf Flowchart of the MC-MCTDH algorithm}. Deterministic propagation steps with multi-configuration time-dependent Hartree steps (MCTDH) with stochastic quantum jumps on the wavefunction due to system-reservoir coupling.}
 \label{fig:algorithm}
 \end{figure}

For Markovian open quantum systems \cite{Breuer2002,Carmichael-book1,*Carmichael-book2}, the evolution for system density matrix $\hat \rho_S$ in Liouville space is determined by a quantum master equation, which in Lindblad form reads  ($\hbar\equiv 1$ is used throughout) \cite{Manzano2020}
\begin{equation}
\begin{aligned}
\frac{d}{dt}{\hat\rho_{\mathrm{S}}}(t)=&\ic[\hat{H}_{\mathrm{S}},\hat\rho_{\mathrm{S}}(t)] + \sum_{j} \hat{L}_{j}\hat\rho_{\mathrm{S}}(t)\hat{L}_{j}^{\dagger} - \frac{1}{2}\{\hat{L}_{j}^{\dagger}\hat{L}_{j},\hat\rho_{\mathrm{S}}(t)\}
\end{aligned}
\label{eq:mastereq}
\end{equation}
where $\hat{H}_{\mathrm{S}}$ is the system Hamiltonian and $\hat{L}_{j}$ are Lindblad {jump operators}  that describe the interaction between the system and the $j$-th reservoir channel. The square brackets denote the commutator and curly brackets   the anticommutator. The Lindblad form of the master equation is a dynamical semi-group that ensures the positivity of the density matrix \cite{Manzano2020}.
The Monte Carlo wavefunction technique avoids the direct integration of the quantum master equation by propagating an initial wavefunction $\Psi(0)$ over a sequence of non-Hermitian evolution intervals that are interrupted at random times by quantum jumps that encode the physics of the Lindblad operators $\hat L_j$ \cite{*Molmer93}.

Figure \ref{fig:algorithm} summarizes the proposed Monte Carlo MCTDH (MC-MCTDH) algorithm. Starting from a reference time $t$, the wavefunction $|\Psi(t)\rangle$ is propagated with the MCTDH method up to $t+\Delta t$ with the effective non-Hermitian Hamiltonian 
\begin{equation}
\hat{H}=\hat{H}_{\mathrm{S}} - \frac{\ic}{2}\sum_{j}\hat{L}^{\dagger}_{j}\hat{L}_{j}.
\label{eq:nonHermitian}
\end{equation}
For small $\Delta t$, the wavefunction norm is reduced as
\begin{equation}
\langle \Psi(t+\Delta t) | \Psi(t+\Delta t) \rangle = 1 - \delta p,
\end{equation}
where $\delta p = \sum_{j}\delta p_{j}$ is determined by the instantaneous jump probabilities $\delta p_{j}=\Delta t\average{\Psi(t)}{\hat{L}_{j}^{\dagger}\hat{L}_{j}}{\Psi(t)}$. At the end of the interval, a pseudo-random number $0<\epsilon<1$ is generated from a uniform distribution, and compared with  $\delta p$. If $\delta p \leq  \epsilon$, no quantum jump occurs and the wavefunction is renormalized as 
\begin{equation}
|\Psi'(t+\Delta t)\rangle = \frac{|\Psi(t+\Delta t)\rangle}{\sqrt{1-\delta p}},
\label{eq:wfnormalized}
\end{equation}
before another interval begins. If $\delta p > \epsilon$ a quantum jump occurs and a Lindblad jump operator is chosen to act on the wavefunction. The $j$-th reservoir channel is chosen such that the operator $\hat L_j$ gives the smallest jump probability $\delta_j$ that is greater than $\epsilon$. The new wavefunction after the jump becomes 
\begin{equation}
|\Psi'(t+\Delta t)\rangle = \frac{\hat{L}_{j}|\Psi(t)\rangle}{\sqrt{\delta p_{j}/\Delta t}},
\end{equation}
and a new interval begins. This algorithm (see Fig. \ref{fig:algorithm}) is sequentially repeated until the propagation ends, resulting in a piecewise quantum trajectory for the system wavefunction. 

Observables are computed by averaging instantaneous expectation values over multiple trajectories \cite{*Molmer93}. For the $k$-th quantum trajectory, expectation values $\langle\Psi^{(k)}(t)|\hat{O}|\Psi^{(k)}(t)\rangle$ are computed at the end of each interval, after normalizing the wavefunction. The procedure is straightforward to extend for computing two-time correlation functions \cite{Breuer2002}. By construction, any trajectory-averaged observable
\begin{equation}
\overline{\langle\hat{O}(t)\rangle} = \frac{1}{n_{\mathrm{T}}}\sum_{k=1}^{n_{\mathrm{T}}}\langle\Psi^{(k)}(t)|\hat{O}|\Psi^{(k)}(t)\rangle
\label{eq:exvalop}
\end{equation}
asymptotically converges to the density matrix solution $\langle\hat{O}\rangle \equiv \mathrm{Tr}[\hat\rho_{\mathrm{S}}(t)\hat{O}]$ with increasing number of trajectories $n_{\rm T}$. The convergence proof can be found in Ref. \cite{*Molmer93} and is reproduced in Appendix \ref{app:lindblad}. In practice, quantum optics problems allow for $n_{\rm T}\sim 10^2$. Mean-square errors (MSE) of the average observables can be defined by comparing with the exact density matrix solutions (see Appendix \ref{app:lindblad}).  The independence of each trajectory facilitates computational parallelization strategies.

\subsection{\label{sec:mctdh}MCTDH Propagator}

We use the MCTDH method for the deterministic propagation step in Fig. \ref{fig:algorithm}. The method was developed by Meyer, Manthe and Cederbaum in 1990 \cite{Meyer1990} as a generalization of the time-dependent Hartree ansatz \cite{TDH1964} for solving the time-dependent Schr\"odinger equation. MCTDH is widely used in chemical physics due to its ability for obtaining essentially exact fully-quantum results for complex molecular systems with a large number of vibrational modes and strong non-adiabatic interactions \cite{Raab99,MCTDHBook2009}.

The standard ansatz for solving the time-dependent Schr\"odinger equation is an expansion in a time-independent basis with time-dependent coefficients of the form
\begin{equation}\label{eq:static ansatz}
\boldsymbol{\Psi}(q_{1},...,q_{f},t)=\sum_{j_{1}=1}^{N_{1}}\dots \sum_{j_{f}=1}^{N_{f}} C_{j_{1}...j_{f}}(t) \prod_{k=1}^{f}\chi_{j_{k}}^{(k)}(q_{k}) ,
\end{equation}
where $q_{k}$ are system coordinates, $C_{j_{1}...j_{f}}(t)$ are dynamical expansion coefficients and $\chi^{(k)}_{j_{k}}$ are the time-independent basis functions that describe the $k$-th degree of freedom. For example, in molecular vibration problems there would be $f$ degrees of freedom (e.g., vibrational modes)  in this expansion, each described by complete basis of $N_{j_k}$ basis functions (e.g., vibrational eigenfunctions) represented on a one-dimensional coordinate grid using discrete variable representation (DVR) techniques \cite{Light2000}.

MCTDH generalizes the static product basis in Eq. (\ref{eq:static ansatz}) with a linear combination of time-dependent Hartree products of the form 
\begin{eqnarray}
\nonumber \boldsymbol{\Psi}(q_{1},...,q_{f},t)&=&\sum_{j_{1}=1}^{n_{1}}\dots \sum_{j_{f}=1}^{n_{f}} A_{j_{1}...j_{f}}(t) \prod_{k=1}^{f}\phi_{j_{k}}^{(k)}(q_{k},t)  \\
\label{eq:mctdh}&=& \sum_{J}\boldsymbol{A}_{J}(t)\boldsymbol{\Phi}_{J}(t)
\end{eqnarray}
where the collective index $J$ labels the set of basis functions $\phi_{j_k}$ in a given tensor product configuration that contributes to the wavefunction, the tensor $\boldsymbol{A}_{J}(t)\equiv A_{j_{1}...j_{f}}(t)$ contains the time-dependent amplitudes of each product configuration, and $\boldsymbol{\Phi}_{J}(t)\equiv \prod_{k=1}^{f}\phi_{j_{k}}^{(k)}(q_{k},t)$ denotes the instantaneous product basis configurations.
The number of relevant basis states per configuration and degree of freedom $n_{k}$ in Eq. (\ref{eq:mctdh}) is typically smaller than the number of DVR basis functions $N_{k}$ needed for convergence the static ansatz in Eq. (\ref{eq:static ansatz}). As a result of this dynamical Hilbert space contraction, the number of product configurations $n_{1}\times n_{2}\times\dots\times n_{f}$ needed for convergence is usually smaller than the number of static configurations in the standard method because $n_k<N_k$ for each of the $k$-th degrees of freedom, which becomes  important when solving high-dimensional quantum dynamics problems. 

The time-dependent Schr\"odinger equation is solved with the MCTDH ansatz [Eq. (\ref{eq:mctdh})] and the Dirac-Frenkel variational principle \cite{MCTDHBook2009}. Coupled non-linear equations for the  $\mathbf{A}_J$ and $\mathbf{\Phi}_J$ tensors are usually derived by introducing projectors over individual degrees of freedom $\hat P^{(k)}\equiv\sum_{j=1}^{n_{k}} | \phi_{j}^{(k)} \rangle \langle \phi_{j}^{(k)} |$,  are complete in the limit $n_k\rightarrow \infty$.  Projecting  Eq. (\ref{eq:mctdh}) over the $k$-th degree of freedom gives 
\begin{equation}
\hat P^{(k)}\boldsymbol{\Psi}=\sum_{l=1}^{n_{k}} | \phi_{l}^{(k)} \rangle \langle \phi_{l}^{(k)} | \boldsymbol{\Psi}\rangle_{k} = \sum_{l=1}^{n_{k}} \phi_{l}^{(k)}\boldsymbol{\Psi}_{l}^{(k)},
\label{eq:mctdhr}
\end{equation}
which for the $k=1$ degree-of-freedom, for example, would give an expansion in the complementary space of the form $\boldsymbol{\Psi}_{l}^{(1)}=\sum_{j_{2}}^{n_{2}}\cdots\sum_{j_{f}}^{n_{f}}A_{l j_{2}\cdots j_{f}}(t)\phi_{j_{2}}^{(2)}\cdots\phi_{j_{f}}^{(f)}$. Variations of the time-dependent coefficients $\mathbf{A}_{J}$ and one-dimensional time-dependent functions $\boldsymbol{\Phi}_{J}$ are given by 
\begin{align}
\label{eq:diracA}\frac{\delta\Psi}{\delta\mathbf{A}_{J}}&=\mathbf{\Phi}_{J} \\
\label{eq:diracpsi}\frac{\delta\Psi}{\delta\phi_{j_{k}}^{(k)}}&=\mathbf{\Psi}_{j_{k}}^{(k)} \\
\label{eq:diract}\dot{\mathbf{\Psi}}=\sum_{J}\dot{\mathbf{A}}_{J}\mathbf{\Phi}_{J} &+ \sum_{k=1}^{f}\sum_{j_{k}=1}^{n_{k}}\dot{\phi}_{j_{k}}^{(k)}\mathbf{\Psi}_{j_{k}}^{(k)}.
\end{align}
Equations of motion for tensor coefficients $\mathbf{A}(t)$ are derived  using Eqs. (\ref{eq:mctdh}), (\ref{eq:diracA}) and (\ref{eq:diract}) in the Dirac-Frenkel variational principle $\langle\delta\boldsymbol{\Psi}| \hat{H} | \boldsymbol{\Psi}\rangle = \ic\left\langle\delta\boldsymbol{\Psi}\left|\frac{\partial}{\partial t} \right| \boldsymbol{\Psi} \right\rangle$ to get the set of coupled nonlinear equations
\begin{align}
\sum_{L}\boldsymbol{A}_{L} \left\langle \boldsymbol{\Phi}_{J} \left| \hat{H} \right| \boldsymbol{\Phi}_{L} \right\rangle =& \ic\sum_{L}\dot{\boldsymbol{A}}_{L} \left\langle \boldsymbol{\Phi}_{J} |  \boldsymbol{\Phi}_{L} \right\rangle \\
\nonumber +& \ic\sum_{k=1}^{f}\sum_{l=1}^{n_{k}}  \left\langle \boldsymbol{\Phi}_{J} |  \dot{\phi}_{l}^{(k)}\boldsymbol{\Phi}_{L} \right\rangle
\end{align}
and
\begin{align}
\ic\dot{\boldsymbol{A}}_{J} =& \sum_{L}\boldsymbol{A}_{L} \left\langle \boldsymbol{\Phi}_{J} \left| \hat{H} \right| \boldsymbol{\Phi}_{L} \right\rangle 
- \ic\sum_{k=1}^{f}\sum_{l=1}^{n_{k}} \boldsymbol{A}_{J^{k}_{l}} g^{(k)}_{jl}
\label{eq:coeffA}
\end{align}
with the constraint $g^{(k)}_{jl}=\ic\langle \phi_{j}^{(k)}|\dot{\phi}_{l}^{(k)}\rangle=\ic\langle \phi_{j}^{(k)}|\hat{g}^{(k)}|\phi_{l}^{(k)}\rangle$. In Eq. (\ref{eq:coeffA}), the tensor elements per degree of freedom are denoted as $\boldsymbol{A}_{J^{k}_{l}}\equiv A_{j_{1} \cdots l \cdots j_{f}}$. 

The equations of motion for the functions $\phi_{j_{k}}^{(k)}$ are also found variationally from the time-dependent Schr\"{o}dinger equation. In terms of the projection operators $\hat P^{(k)}$ they read 
\begin{align}
&\sum_{l_{k}}\langle\hat{H}\rangle_{j_{k}l_{k}}^{(k)}\phi_{l_{k}}^{(k)} = P^{(k)}\sum_{l_{k}}\langle\hat{H}\rangle_{j_{k}l_{k}}^{(k)}\phi_{l_{k}}^{(k)} + \ic\sum_{l_{k}}\rho_{j_{k}l_{k}}^{(k)}\dot{\phi}_{l_{k}}^{(k)} \\
&\ic\dot{\phi}_{j_{k}}^{(k)}=\sum_{l_{k},m_{k}}\left(\boldsymbol{\rho}^{(k)^{-1}}\right)_{j_{k}l_{k}}\left( 1- P^{(k)}\right)\langle\hat{H}\rangle_{l_{k}m_{k}}^{(k)}\phi_{m_{k}}^{(k)},
\label{eq:onefuncs}
\end{align}
where $\rho^{(k)}_{j_{k}l_{k}}\equiv \langle\boldsymbol{\Psi}^{(k)}_{j_{k}}|\boldsymbol{\Psi}^{(k)}_{l_{k}}\rangle$ is a reduced density matrix, and $\langle\hat{H}\rangle_{jl}^{(k)}=\langle\boldsymbol{\Psi}_{j}^{(k)} | \hat{H} | \boldsymbol{\Psi}_{l}^{(k)}\rangle$  the mean-field Hamiltonian of the $k$-th degree of freedom.  The solution of the MCTDH equations of motion  preserve the norm and total energy for time-independent Hermitian Hamiltonians. In this work we use an extension of the method that includes non-Hermitian (complex) potentials. Additional details about  MCTDH can be found in Refs. \cite{MCTDHBook2009,Beck2000,mctdhintro}. 

\section{\label{sec:results} Results}

We test the MC-MCTDH methodology by solving selected open quantum system problems of interest in cavity QED. For comparison, we also solve the corresponding Lindblad quantum master equation for the density matrix using the open source Python library QuTiP \cite{JOHANSSON2012}. The same desktop machine is used for all calculations (CPU 3 GHz Intel Core i5, 8 GB RAM), unless otherwise stated.

\subsection{\label{sec:cavlosses} Cavity field with finite photon lifetime}

 \begin{figure}[!t]
\includegraphics[width=0.4\textwidth]{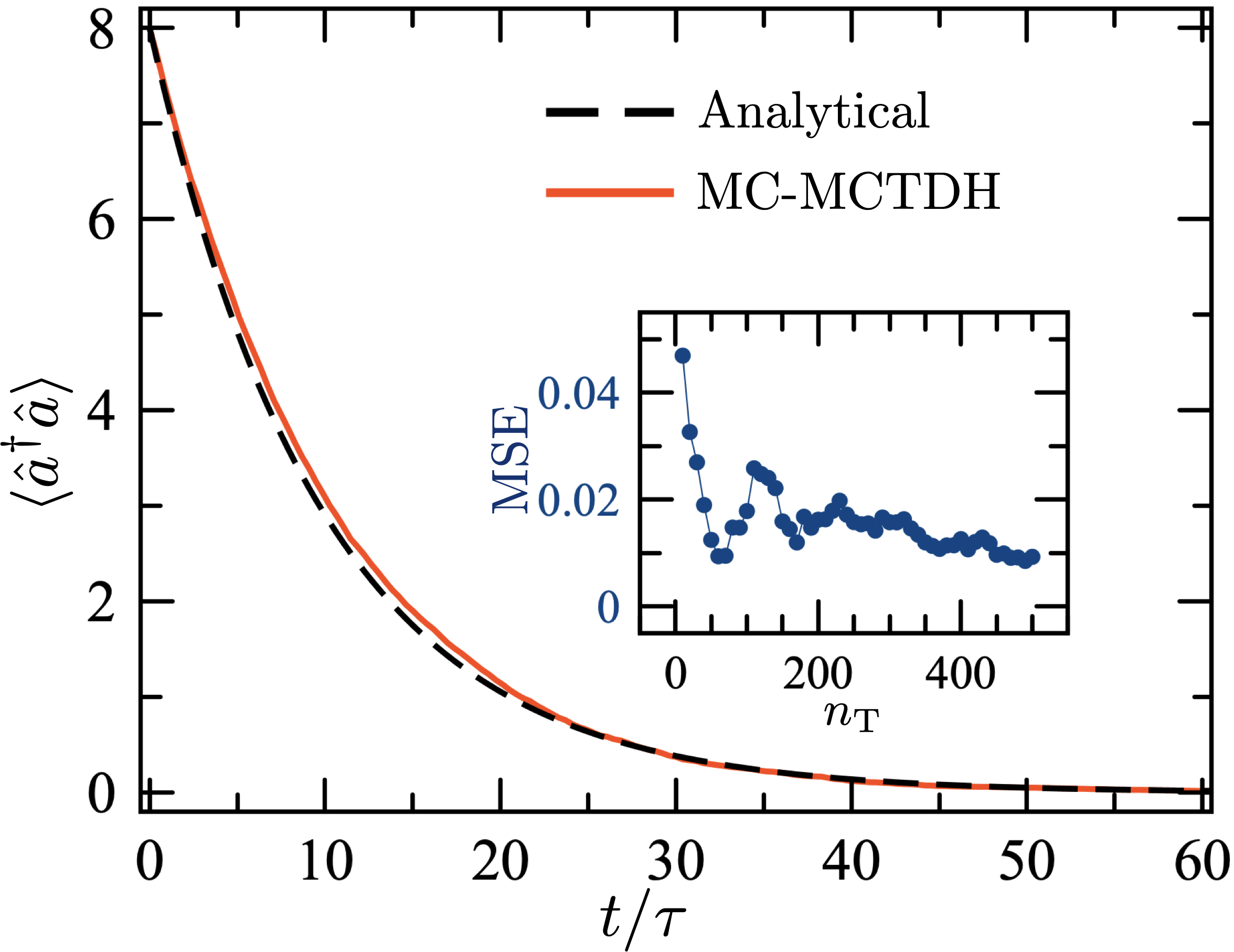}
 \caption{\textbf{Exponential decay of a lossy cavity.}
Simulated decay of an $n=8$ Fock state with $n_{\mathrm{T}}=200$  MC-MCTDH quantum trajectories (solid line). The analytical solution is shown for comparison (dashed line). Time is in units of the cavity oscillation period $\tau=2\pi/\omega_{\rm c}$ and $\kappa=0.016\,\omega_{\mathrm{c}}$ is the photon decay rate. Inset: Mean-squared error (MSE) as a function of the number of trajectories.}
 \label{fig:cavity}
 \end{figure}

Consider a cavity mode with resonance frequency $\omega_{\rm c}$ in a  structure with imperfect mirror reflectivity. The mode is modeled as a quantum harmonic oscillator with annihilation operator $\hat a$. Cavity photons  leak out to the far field at rate $\kappa$. A minimal decoherence model for radiative decay can be constructed with the Lindblad operator $\hat L_\kappa=\sqrt{\kappa}\,\hat{a}$. The effective Hamiltonian for the deterministic steps of the Monte Carlo propagation method is thus given by
\begin{equation}
\hat{H}=(\omega_{\mathrm{c}}-{\ic\kappa}/{2})\,\hat{a}^{\dagger}\hat{a}.
\end{equation}

The Heisenberg equation of motion for the number operator $\hat n=\hat a^\dagger\hat a$ has the simple solution $\langle\hat n (t)\rangle= \langle\hat{n}(0)\rangle\,{\rm exp}[-\kappa t]$, i.e., exponential decay of initial occupation number $\meanval{\hat{n}(0)}$ with population decay time $T_1=1/\kappa$. In Fig. \ref{fig:cavity} we show the MC-MCTDH evolution of an initial Fock state with  $\meanval{\hat{n}(0)}=8$ photons, together with the analytical solution. The inset shows that ${\rm MSE}\approx 1\%$ can be achieved with about $n_{\rm T}=400$ quantum trajectories.

This one-dimensional example is not exploiting the MCTDH tensor ansatz, but demonstrates the stochastic quantum jumps on a  DVR grid for an excited Fock state. In general, most photonic states of interest in quantum optics (Fock state, coherent states, squeezed light) can be accurately represented with DVR grids \cite{Triana2018}, which could be advantageous in comparison with more elaborate phase-space representations \cite{Zhu2019,Koessler2022}.

\subsection{\label{sec:rabi} Vacuum Rabi oscillations}

Our next case study is two bilinearly coupled quantum harmonic oscillators in the rotating wave approximation. For an initial state with a single excitation in one of the oscillators, i.e., $\ket{\Psi(0)}=\ket{1}\ket{0}$, we expect the MC-MCTDH algorithm to describe damped Rabi oscillations of the subsystem variables. For a harmonic oscillator with annihilation operator $\hat b$ and resonance frequency $\omega_{0}$ (e.g., molecular vibration) interacting with the a cavity field $\hat a$ of frequency $\omega_c$, the effective Hamiltonian is given by
\begin{equation}
\hat{H} = (\omega_{\mathrm{c}}-i\kappa/2)\hat{a}^{\dagger}\hat{a} + (\omega_{0}-i\gamma/2)\hat{b}^{\dagger}\hat{b} + g(\hat{b}^{\dagger}\hat{a} + \hat{b}\hat{a}^{\dagger}),
\end{equation}
where $g$ is the coupling strength (Rabi frequency). Dissipation of the $ b$-oscillator at  rate $\gamma$ is modeled with the Lindblad operator $\hat L_\gamma= \sqrt{\gamma}\,\hat b$, and cavity dissipation is described as before. 

\begin{figure}[!t]
 \includegraphics[width=0.4\textwidth]{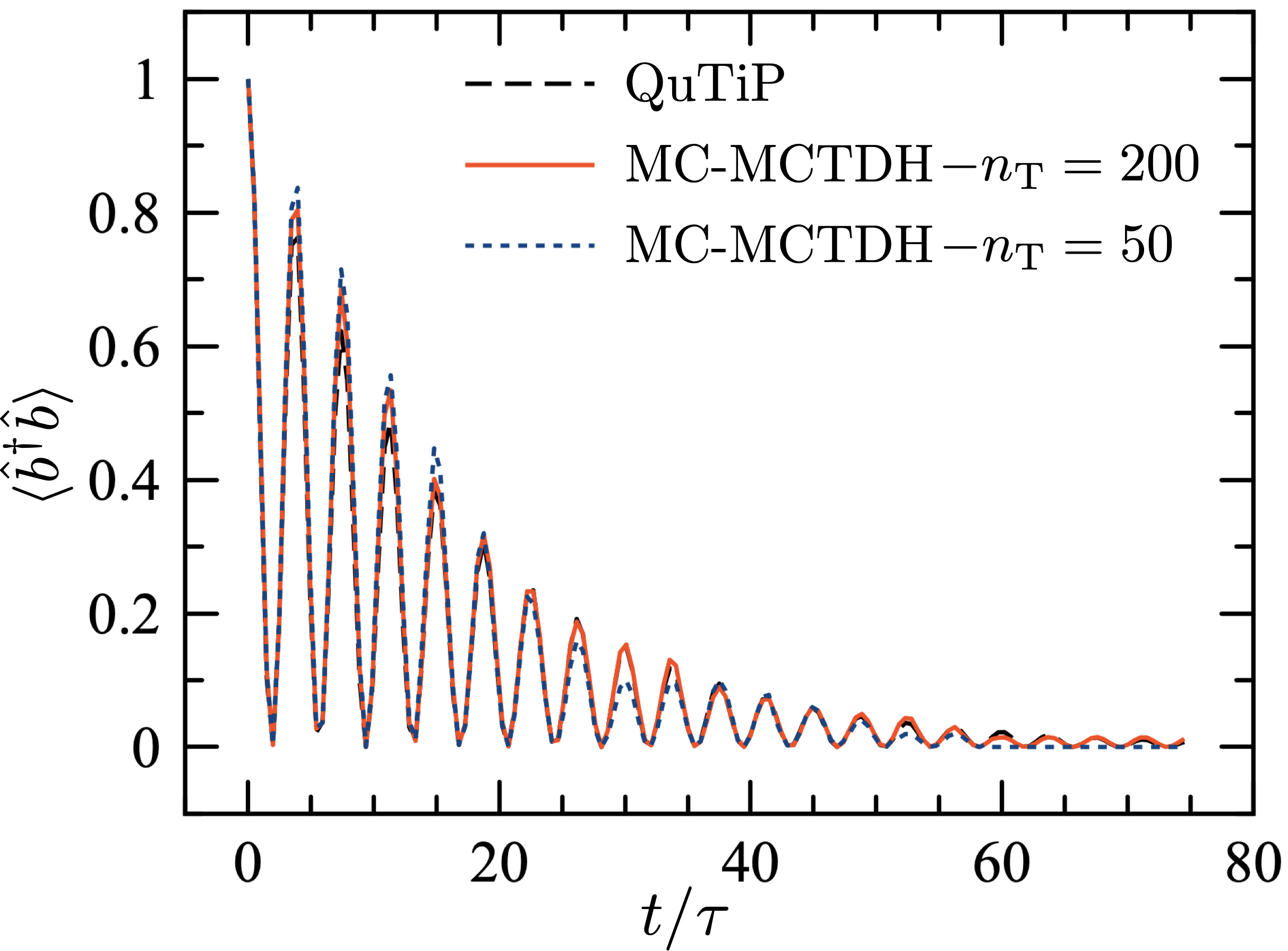}
 \caption{\textbf{Vacuum Rabi oscillations.} Coherent transfer of a single initial excitation between two resonantly coupled harmonic oscillators, obtained with MC-MCTDH for two sets of quantum trajectories $n_T=(50,200)$. The QuTiP Liouville space solution is show for comparison (black dashed line). The frequency of the $b$-oscillators is $\omega_{0}$, the dissipation rates are $\kappa=0.026\omega_{0}$ and $\gamma=0.013\omega_{0}$ and the bilinear coupling strength is $g=0.13\omega_{0}$. Time in units of $\tau\equiv 2\pi/\omega_{0}$.
}
 \label{fig:twolevel}
 \end{figure}

Figure \ref{fig:twolevel} shows the evolution of the occupation number $\langle \hat b^\dagger \hat b\rangle$ obtained with MC-MCTDH for strong resonant coupling ($\omega_{\rm c}=\omega_0$ and $g/\omega_{\rm c}>0.1$). The exact density matrix solution is also shown for comparison. For the chosen system parameters,  averaging over $n_{\mathrm{T}}=50$ quantum trajectories gives good short-time accuracy, although errors tend to accumulate at longer times ($t>20\times 2\pi/\omega_{\rm c}$). Increasing the number of trajectories ($n_{\mathrm{T}}\sim 200$) gives results that match better the exact Liouville-space solution. 

\subsection{\label{sec:revivals} Jaynes-Cummings revivals in driven cavities}

\begin{figure}[!t]
\includegraphics[width=0.4\textwidth]{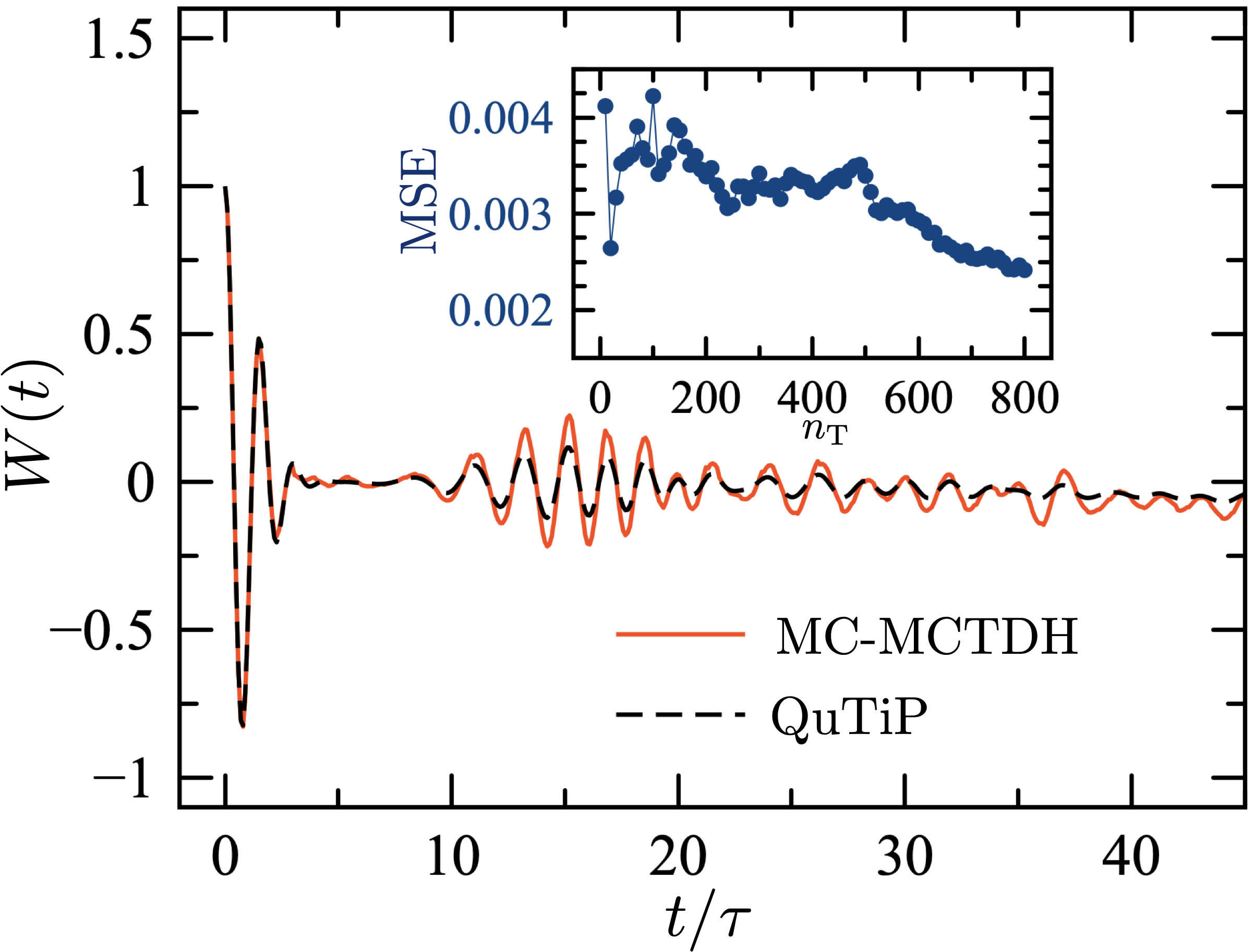} \\
\caption{\textbf{Jaynes-Cummings population revivals in a driven cavity.} MC-MCTDH evolution of the atomic inversion $W(t)$ for a qubit in a cavity initially prepared in a coherent state with $|\alpha|=5$ average photons, for $n_{\mathrm{T}}=400$ quantum trajectories (solid red line). The Liouville-space solution is shown for comparison (dashed black line). Inset: Mean square error (MSE) as a function of the number of trajectories. The qubit frequency is $\omega_{0}$, the dissipation rates are $\kappa=\gamma=3.5\times10^{-3}\omega_{0}$ and the Rabi frequency is $g=0.13\omega_{0}$. Time in units of $\tau=2\pi/\omega_0$. 
}
\label{fig:qubit}
\end{figure}

We now study the coupling of a two-level atom (qubit) with a cavity field $\hat a$ prepared in a coherent state $\ket{\alpha}$ with (real) amplitude $\alpha$. In the number (Fock) basis, the coherent state gives a Poissonian distribution with $\langle \hat a^\dagger \hat a \rangle=|\alpha|^2$ \cite{Carmichael-book1}. Unitary dynamics is governed by the Jaynes-Cummings Hamiltonian \cite{Jaynes1963,Gerry2005}
\begin{align}
\hat{H}_{0} = \frac{1}{2}\omega_{0}\hat{\sigma}_{z} + \omega_{\mathrm{c}}\hat{a}^{\dagger}\hat{a} +  g(\hat{\sigma}_{+}\hat{a} + \hat{\sigma}_{-}\hat{a}^{\dagger}), 
\label{eq:jcham}
\end{align}
$\hat{\sigma}_{z}$  and $\hat{\sigma}_{\pm}$ are the Pauli spin-$1/2$ operators, $\omega_0$ is the qubit frequency, and $g$ the Rabi frequency. Since spins do not have a coordinate dependence, we represent the two spin projections $m_z=\pm 1/2$ in a DVR grid using two flat potentials separated in energy by $\omega_0$, i.e., $V_\pm(q)=\pm \,\omega_0/2$. This is equivalent to having two electronic states in the MCTDH package \cite{MeyerMCTDH2007}. Pauli  operators can be constructed accordingly. Atomic relaxation at the rate $\gamma$ is given by the Lindblad operator $\hat L_\gamma=\sqrt{\gamma}\,\hat \sigma_-$, and cavity decay is described before. The effective Hamiltonian for deterministic propagation is given by $\hat H= \hat H_0 - i(\kappa/2)\, \hat a^\dagger \hat a- i(\gamma/2)\, \hat \sigma_{+} \hat\sigma_{-}$, with $\hat H_0$ in Eq. (\ref{eq:jcham}). 

Figure \ref{fig:qubit} shows the evolution of the inversion $W(t) = \rho_{ee}(t)- \rho_{gg}(t)$, where $\rho_{ii}$ denotes level population. The qubit is initialized in the excited level ($W(0)=1$) and strongly couples to a cavity that has initially $|\alpha|^2=5$ photons on average. The Rabi frequency is $g=0.13\,/\omega_0$. For the small damping rates used ($\kappa=\gamma\sim 10^{-3}\omega_0$), the MC-MCTDH reproduces the long-time revivals of the population inversion expected due to the exchange of coherence between qubit and Fock sub-levels that compose the coherent state \cite{Gerry2005}, using only $n_{\rm T}=400$ quantum trajectories. Deviations from the exact Liouville-space solution are negligible at short times, but grow over longer timescales. The inset in Fig. \ref{fig:qubit} shows the drop of the MSE with increasing number of  trajectories.

\subsection{\label{sec:nqho} Independent quantum oscillators coupled to a common cavity field}

In this example, we consider a set of $N$ independent oscillators $\hat b_i$ that couple with a common quantized cavity field $\hat a$. The non-Hermitian effective Hamiltonian of the system is given by
\begin{equation}
\begin{aligned}
\hat{H} &= \omega_{\mathrm{c}}\hat{a}^{\dagger}\hat{a}  - \frac{\ic\kappa}{2}\hat{a}^{\dagger}\hat{a} \\
&  + \sum_{i=1}^{N}\left[\omega_{0}\hat{b}_{i}^{\dagger}\hat{b}_{i} + g(\hat{b}_{i}^{\dagger}\hat{a} +\hat{b}_{i}\hat{a}^{\dagger}) - \frac{\ic\gamma}{2}\hat{b}_{i}^{\dagger}\hat{b}_{i} \right], 
\end{aligned}
\end{equation}
with jump operators for the $b$-oscillators $\hat{L}_{i\gamma} = \sqrt{\gamma}\hat{b}_{i}$, and cavity dissipation described as before. We focus on the coherent population transfer between oscillators beyond the single-excitation manifold.

\begin{figure}[!t]
\includegraphics[width=0.4\textwidth]{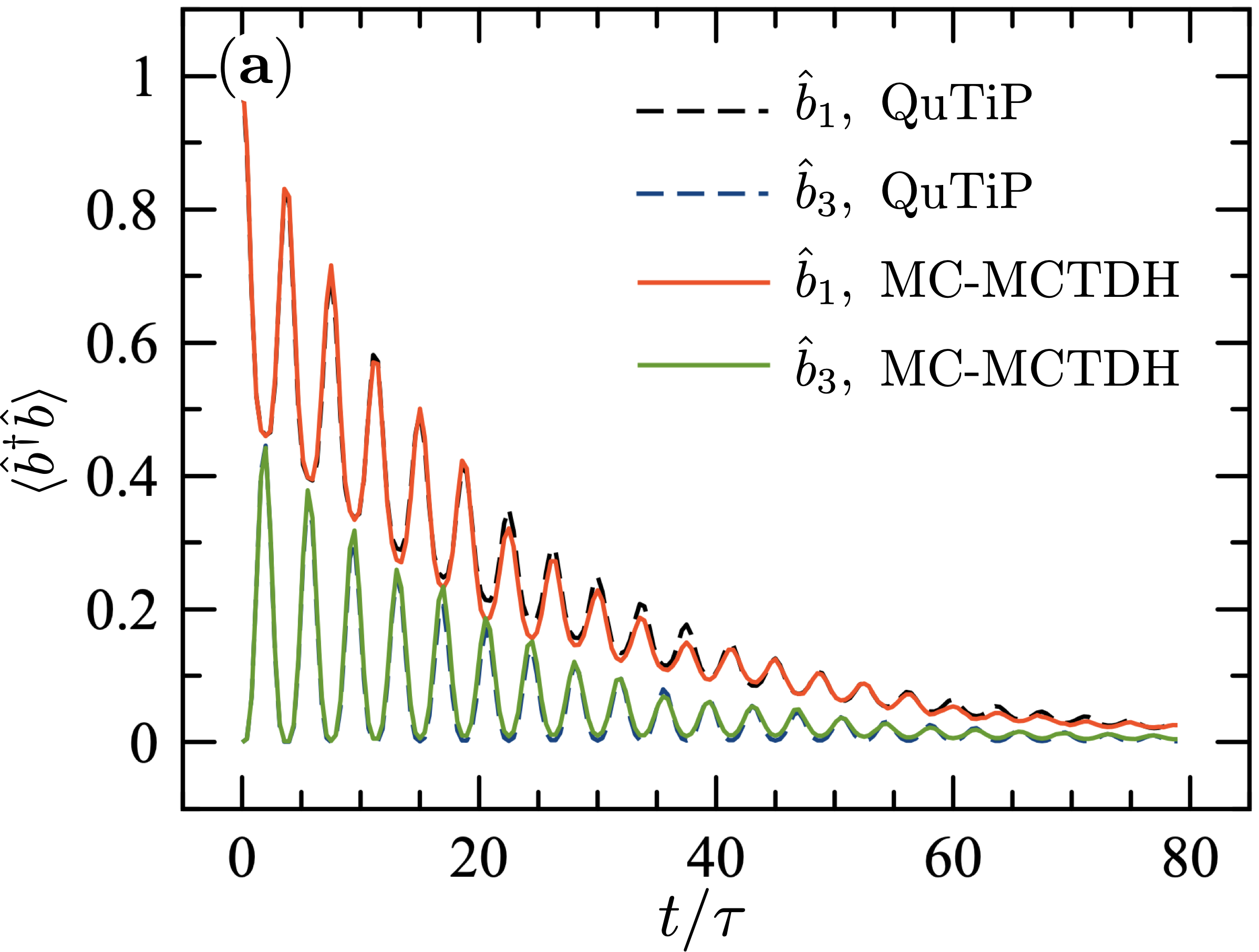} \\
\includegraphics[width=0.4\textwidth]{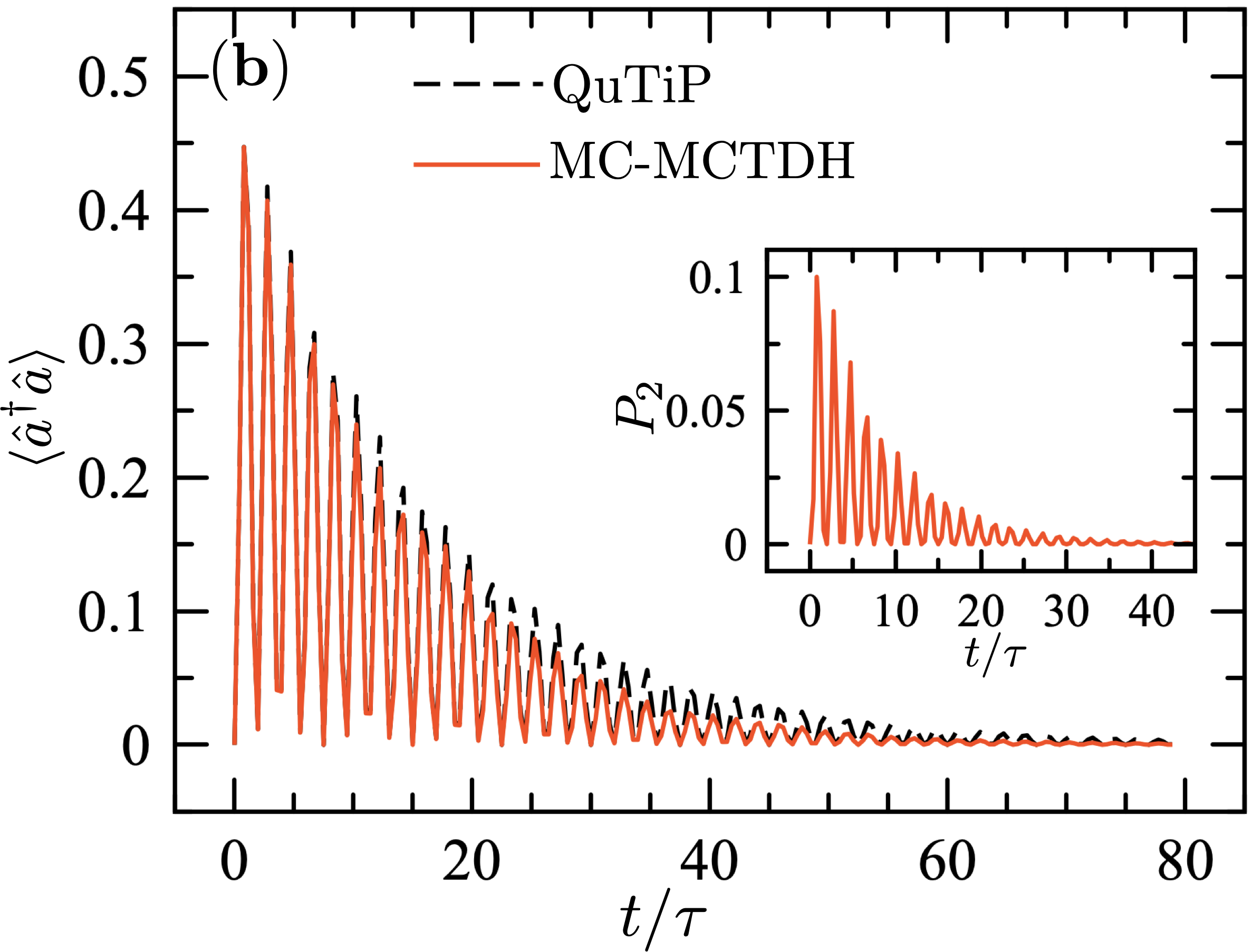}
\caption{\textbf{Cavity-mediated population transfer between $N$ uncoupled oscillators.} (a) MC-MCTDH energy transfer between subsystems ($b_1, b_3$) in a set of $N=4$ oscillators coupled resonantly with a cavity field, with $n_{\mathrm{T}}=200$ quantum trajectories (solid lines). (b) Photon occupation number $\langle \hat a^\dagger \hat a\rangle$ of cavity field using MC-MCTDH (solid red line) and solving the Lindblad master equation (black dashed line). Inset: Time-dependent projection over Fock state with $n=2$.
The frequencies of the $b$-oscillators are $\omega_{0}$, the dissipation rates are $\kappa=0.026\omega_{0}$ and $\gamma=0.013\omega_{0}$, and the bilinear coupling strength is $g=0.13\omega_{0}$.  Time in units of $\tau=2\pi/\omega_0$.
}
\label{fig:ntwolevel}
\end{figure}

Figure \ref{fig:ntwolevel}a shows the MC-MCTDH evolution of the occupation numbers $\langle \hat b_1^\dagger \hat b_1\rangle$ and $\langle \hat b_3^\dagger \hat b_3\rangle$ for a set of $N=4$ oscillators in a cavity that initially has one excitation in $b_1$ and another excitation in $b_2$, with the cavity field in the vacuum, i.e., $|\Psi(0)\rangle=|1,1,0,0,n=0\rangle$. The results converge to the exact Liouville-space solution for the $b$-variables for $n_T=200$ quantum trajectories. However, Fig. \ref{fig:ntwolevel}b shows that long-time errors of about $4\%$ tend to accumulate for the photon number $\langle \hat a^\dagger \hat a\rangle$ after several population transfer cycles, which could be reduced by increasing $n_T$. The inset of \ref{fig:ntwolevel}b shows that the method captures the short-time rise and long-time decay of the $n=2$ Fock state population $P_2$. Since the amount of ground state bleaching is significant ($>10\%$), the Hilbert space dimension needed to converge the Liouville-space solution is higher than the previous examples. For the parameters in Fig. \ref{fig:ntwolevel}, QuTiP solutions converge with a minimum Hilbert space dimension of $d \equiv (\nu_{\rm max}+1)^N \times (n_{\rm max}+1)=324$, where $\nu_{\rm max}=2$ is the maximum quantum number used for each of the $b$-oscillators and $n_{\rm max}=3$ is the highest Fock state included.

\subsection{\label{sec:nqhoc} Strongly interacting array of quantum oscillators coupled to a common cavity field}

As a final example, we compute the dynamics of a circular array of quantum harmonic oscillators of size $N$ with periodic boundary conditions. The array oscillator couple strongly to a common cavity field. We monitor the population transfer dynamics between oscillators in the array with the MC-MCTDH method assuming strong bilinear coupling between sites. The  effective Hamiltonian is given by
\begin{eqnarray}
\nonumber \hat{H} &=& \omega_{\mathrm{c}}\hat{a}^{\dagger}\hat{a}  - \frac{\ic\kappa}{2}\hat{a}^{\dagger}\hat{a} \\
\label{eq:ncouosc} & & + \sum_{i=1}^{N}\left[\omega_{0}\hat{b}_{i}^{\dagger}\hat{b}_{i} + g(\hat{b}_{i}^{\dagger}\hat{a} +\hat{b}_{i}\hat{a}^{\dagger}) - \frac{\ic\gamma}{2}\hat{b}_{i}^{\dagger}\hat{b}_{i} \right] \\
\nonumber & & + \sum_{i=1}^{N-1} \lambda(\hat{b}^{\dagger}_{i}\hat{b}_{i+1} + \hat{b}_{i}\hat{b}_{i+1}^{\dagger}) + \lambda(\hat{b}^{\dagger}_{1}\hat{b}_{N} + \hat{b}_{1}\hat{b}_{N}^{\dagger})
\end{eqnarray}
where $\lambda$ is the nearest-neighbor coupling and the other variables are defined as before. 

\begin{figure}[!t]
\includegraphics[width=0.4\textwidth]{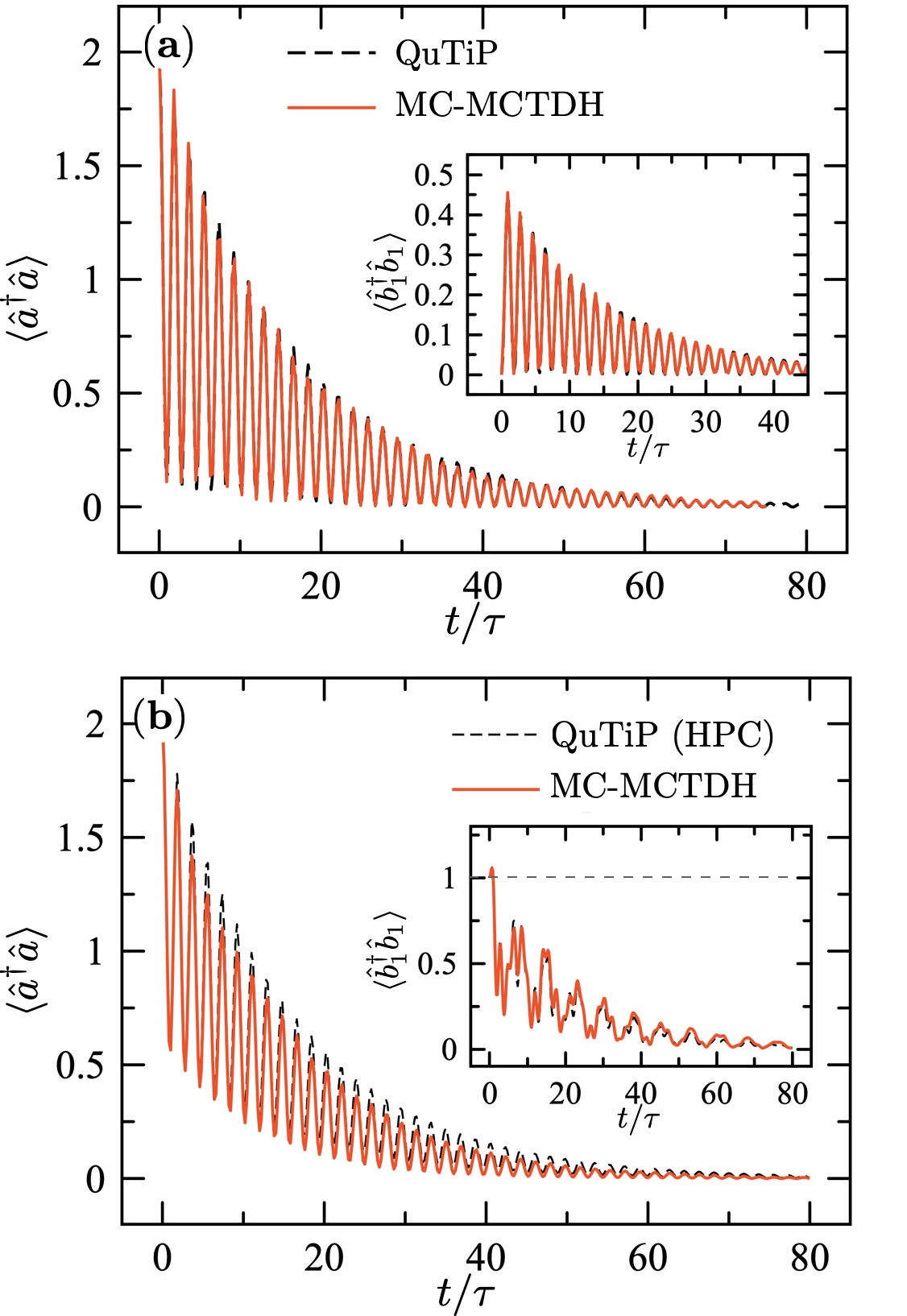}
\caption{\textbf{Population transfer for a strongly-coupled oscillator array in a cavity.} 
(a) MC-MCTDH evolution of the cavity occupation number $\langle\hat{a}^{\dagger}\hat{a}\rangle$ for an array of $N=4$ oscillators initially in the ground state inside a cavity with $n=2$ photons (solid red line), for $n_{\rm T}=300$ quantum trajectories. The Lioville-space solution (QuTiP) is also shown (dashed black line); (b) Same as panel (a) for two initial array excitations and two cavity photons. The density matrix solution  was obtained with an HCP workstation. In both panels, the inset shows the evolution of the occupation number for oscillator $b_1$. The $b$-oscillator frequencies are $\omega_{0}$, the dissipation rates are $\kappa=0.026\omega_{0}$ and $\gamma=0.013\omega_{0}$, the light-matter coupling strength is $g=0.13\omega_{0}$, and the nearest-neighbor coupling in the array is $\lambda=g/2$. Time is in units of $\tau=2\pi/\omega_0$. 
}
\label{fig:nharosc}
\end{figure}

In Fig. \ref{fig:nharosc} we show the occupation numbers of the cavity field and the oscillator $\hat b_1$, for an array of size $N=4$ and an initially excited cavity with $n=2$ photons. Two array excitation levels are studied. In Fig. \ref{fig:nharosc}a the oscillator array is set to the ground state ($\nu_{\rm max}=0$). In this case, the initial cavity excitations are transferred rapidly to the oscillator array creating a many-particle wavepacket that eventually decays within a few vibrational lifetimes. The evolution can be converged in Liouville space with a truncated Hilbert space that includes up to $\nu_{\mathrm{max}}=3$ excitations per site and $n_{\mathrm{max}}=5$ photons, giving the Hilbert space dimension $d=1536$. Converged MC-MCTDH  calculations involved $N_{k}=41$ grid points for each degree of freedom in a harmonic oscillator-DVR primitive basis, with $n_{k}=4$ time-dependent functions, giving 1844 equations of motion to solve. Fig. \ref{fig:nharosc}a shows that the MC-MCTDH expectation values agree with the converged Liouville-space results within $\sim1$\% with only $n_T=300 $ quantum trajectories. 

In Fig. \ref{fig:nharosc}b, we informally probe the efficiency of the MC-MCTDH method by increasing the initial excitation density of the array to two excitations: one excitation in oscillator $b_1$ and another excitation in oscillator $b_2$, again with two initial cavity photons,  i.e., $|\Psi(0)\rangle=|1,1,0,0,n=2\rangle$. For the same Hamiltonian and dissipative parameters in Fig. \ref{fig:nharosc}a, convergence of the Liouville space solution was not possible on the same machine where MC-MCTDH was implemented, due to RAM constraints. We obtained converged density matrix solutions with QuTiP implemented in a high-perfomance computing (HCP) workstation. The  minimum Hilbert space dimension needed for $1\%$ convergence was found to be $d= 10368$, which included $\nu_{\rm max}=5$ excitations per site and $n_{\rm max}=7$ cavity Fock states. 
Fig. \ref{fig:nharosc}b shows that the MC-MCTDH solution obtained in the low-RAM machine agrees well with the numerically-exact Liouville space solutions for $\hat a$ and $\hat b$ oscillators in the HCP workstation, using only 300 quantum trajectories.

\section{\label{sec:conclu}Conclusions and Discussion}

Motivated by current problems in molecular quantum electrodynamics \cite{Feist2018,Herrera2020perspective,GarciaVidal2021}, we developed an efficient numerical methodology for computing the open system dynamics of strongly coupled quantized oscillators. The method combines deterministic non-unitary propagation of the many-particle system wavefunction in coordinate space, with a sequence of stochastic quantum jumps that model the interaction of the system with multiple reservoirs. The stochastic component of the propagator is based on the Monte-Carlo wavefunction method developed in quantum optics \cite{Molmer92}, which by construction converges to Lindblad semi-group dynamics \cite{*Molmer93}. The deterministic steps are implemented using the multi-configuration time-dependent Hartree method (MCTDH \cite{Meyer1990}), which was originally developed to describe wavepackets with continuous-variable degrees of freedom that are relevant in chemical dynamics. 

We demonstrate the applicability of the method by solving the open quantum system dynamics of selected scenarios of current interest: (\emph{i}) decay dynamics of a lossy optical cavity; (\emph{ii}) vacuum Rabi oscillations for strongly interacting cavity-vibration systems with photonic and material losses; (\emph{iii}) population revivals for a two-level system in a driven cavity; (\emph{iv}) photon-mediated population transfer between independent molecular vibrations coupled to a common cavity field; (\emph{v}) quench dynamics in an array of strongly interacting vibrational oscillators with high initial excitation density. In all cases the proposed method converges to the exact Liouville-space solution with a reasonably low number of quantum trajectories. For an array of strongly coupled oscillators with high excitation density, preliminary tests suggest that the method is more efficient than currently available open-source quantum optics libraries \cite{JOHANSSON2012} at equal machine resources.     

Applications of this quantum dynamics methodology include the study of vibrational relaxation and rotational depolarization of molecular ensembles in liquid-phase infrared cavities under vibrational strong coupling \cite{Simpkins2021}, which are believed to determine the dynamics of unconventinoal light-matter coherences that emerge in two-dimensional infrared cavity spectroscopy \cite{Grafton:2021}, and the reactive dynamics of polar molecules under vibrational ultrastrong coupling \cite{Hernandez2019,Triana2020}. The methodology can also be implemented with time-dependent Hamiltonians to study coherent control scenarios in nanophotonics \cite{Muller2018,Metzger2019,Triana2022-ir}. Future extensions of the method can be implemented to describe systems with non-Markovian coupling to multiple reservoirs \cite{Piilo2009}.

\begin{acknowledgments}
We thank Johannes Schachenmayer and Oriol Vendrell for comments. This work was supported by ANID Postdoctoral 3200565,  FONDECYT Regular 1181743, Millennium Science Initiative Program ICN17-012 and Programa de Cooperación Científica ECOS-ANID ECOS200028.
\end{acknowledgments}
    
\section*{Data Availability} 
The data that support the findings of this study are available from the corresponding author upon reasonable request.

\appendix

\section{Equivalence of the Monte Carlo Wavefunction method and Lindblad quantum master equations}
\label{app:lindblad}

The time evolution of wave function $|\Psi(t)\rangle$ inside MCWF is performed by finding the wave function at a time $t+\Delta t$ for enough small $\Delta t$.
At first-order approximation we obtain (in atomic units)
\begin{equation}
|\Psi(t+\Delta t)^{N}\rangle = \left( 1 - \ic\hat{H}\Delta t\right)|\Psi(t)\rangle,
\label{eq:firstorder}
\end{equation}
like $\hat{H}$ is non-Hermitian, $|\Psi(t+\Delta t)^{N}\rangle$ is not normalized and hence
\begin{equation}
\langle\Psi(t+\Delta t)^{N}|\Psi(t+\Delta t)^{N}\rangle = 1 - \delta p,
\end{equation}
with 
\begin{equation}
\delta p=\sum_{n}\delta p_{n} = \Delta t \sum_{n} \average{\Psi(t)}{\hat{L}^{\dagger}_{n}\hat{L}_{n}}{\Psi(t)},
\end{equation}
where $\delta p_{n}$ describes the loss of the norm of jump operator $\hat{L}_{n}$. 

Considering the operator $\hat{\sigma}(t)=|\Psi(t)\rangle\hspace*{-0.5mm}\langle\Psi(t)|$, for a define number of realizations with different random numbers $\epsilon$ at time $t+\Delta t$, the average value of $\hat{\sigma}(t+\Delta t)$ is given by 
\begin{equation}
\begin{aligned}
\overline{\hat{\sigma}(t+\Delta t)} &= (1-\delta p)\ket{\Psi(t+\Delta t)}\hspace*{-1mm}\bra{\Psi(t+\Delta t)} \\
&+ \delta p \sum_{n} \alpha_{n} \ket{\Psi(t+\Delta t)}\hspace*{-1mm}\bra{\Psi(t+\Delta t)},
\end{aligned}
\label{eq:sigaveragedt}
\end{equation}
with $\alpha_{n}=\delta p_{n}/\delta p$.
Inserting Eqs. (\ref{eq:wfnormalized}) and (\ref{eq:firstorder}) into Eq. (\ref{eq:sigaveragedt}), we obtain
\begin{equation}
\begin{aligned}
\overline{\hat{\sigma}(t+\Delta t)} &=  (1-\ic\Delta t\hat{H})\hat{\sigma}(t)(1+\ic\Delta t\hat{H}^{\dagger}) \\
&+ \delta p\sum_{n}\alpha_{n}\frac{\hat{L}_{n}\hat{\sigma}(t)\hat{L}_{n}^{\dagger}}{\delta p_{n}/\delta t}
\end{aligned}
\end{equation}
\begin{equation}
\begin{aligned}
\overline{\hat{\sigma}(t+\Delta t)} &= \hat{\sigma}(t) - \ic\Delta t\hat{H}\hat{\sigma}(t) + \ic\Delta t\hat{\sigma}(t)\hat{H}^{\dagger} \\
& + \Delta t\sum_{n}\hat{L}_{n}\hat{\sigma}(t)\hat{L}_{n}^{\dagger} + \mathcal{O}(\Delta t^{2}),
\end{aligned}
\end{equation}
and considering that $\hat{H}$ is given by Eq. (\ref{eq:nonHermitian}), we obtain
\begin{equation}
\begin{aligned}
\overline{\hat{\sigma}(t+\Delta t)} &= \hat{\sigma}(t) + \ic\Delta t[\hat{\sigma}(t),\hat{H}_{\mathrm{S}}] \\ 
&- \frac{\Delta t}{2}\sum_{n}\{\hat{\sigma}(t),\hat{L}_{n}^{\dagger}\hat{L}_{n}\} \\
& + \Delta t\sum_{n}\hat{L}_{n}\hat{\sigma}(t)\hat{L}_{n}^{\dagger} + \mathcal{O}(\Delta t^{2}).
\end{aligned}
\label{eq:averagef}
\end{equation}

Now, if we apply the limit $\Delta t\to0$, Eq. (\ref{eq:averagef}) reduces to
\begin{equation}
\frac{\mathrm{d}\overline{\hat{\sigma}}}{d\mathrm{t}} = \ic[\overline{\hat{\sigma}},\hat{H}_{\mathrm{S}}] + \mathcal{L}[\overline{\hat{\sigma}}],
\label{eq:mastereqave}
\end{equation}
where $\mathcal{L}[\overline{\hat{\sigma}}]$ is the Lindblad superoperator given by
\begin{equation}
\mathcal{L}[\overline{\hat{\sigma}}] = \sum_{n}\hat{L}_{n}\hat{\sigma}(t)\hat{L}_{n}^{\dagger} - \frac{1}{2}\sum_{n}\{\hat{\sigma}(t),\hat{L}_{n}^{\dagger}\hat{L}_{n}\} .
\end{equation}
Note that Eq. (\ref{eq:mastereqave}) is equivalent to Eq. (\ref{eq:mastereq}). Hence, we demonstrate the validity of MCWF with the master equation in Lindblad form. 

Now, the next step is to calculate the expectation value of a given operator $\hat{O}$, which according to the density operator in the limits $\Delta t\to0$ and $n_{\mathrm{T}}\to\infty$ is equivalent to $\langle\hat{O}\rangle=\mathrm{Tr}[\hat\rho_{\mathrm{S}}(t)\hat{O}]$.
In the MCWF method is calculated by implementing Eq. (\ref{eq:exvalop}).
However, in MCWF there are numerical errors for a finite number of trajectories $n_{\mathrm{T}}$. 
We measure the error by calculating the mean squared error at time $t$ given by 
\begin{equation}
\mathrm{MSE}[\overline{\langle\hat{O}(t)\rangle}]=\frac{1}{n_{\mathrm{T}}}\sum_{k=1}^{n_{\mathrm{T}}}\left[ \langle\hat{O}(t)\rangle_{(k)} - \langle\hat{O}(t)\rangle \right]^{2}
\end{equation}
where $\langle\hat{O}(t)\rangle_{(k)}$ is the expectation value of trajectory $k$ and $\langle\hat{O}(t)\rangle=\mathrm{Tr}[\hat\rho_{\mathrm{S}}(t)\hat{O}]$ is the exact solution.

\bibliographystyle{apsrev4-1}
\bibliography{biblio-jcp}

\end{document}